\documentclass[journal=jctcce, manuscript=article]{achemso}

\usepackage[utf8]{inputenc}

\usepackage[table]{xcolor}
\usepackage[version=4]{mhchem}
\usepackage{amsfonts}
\usepackage{amsmath}
\usepackage{amsthm}
\usepackage{amssymb}
\usepackage{graphicx}
\graphicspath{{img/}}
 \DeclareGraphicsExtensions{.pdf} 
\usepackage{caption}
\usepackage{subcaption}
 \DeclareCaptionSubType*[arabic]{figure}
\usepackage{braket}
\usepackage{multicol}
\usepackage{multirow}
\usepackage{todonotes}
\usepackage{booktabs}
\usepackage[nohyperlinks,nolist]{acronym}
\usepackage{siunitx}
\usepackage{mathtools}
\usepackage{tablefootnote}
\usepackage{multirow}

\begin{acronym}
\acro{AO}{Atomic Orbital}
\acro{MO}{Molecular Orbital}
\acro{API}{Application Programmer Interface}
\acro{AUS}{Advanced User Support}
\acro{BO}{Born-Oppenheimer}  
\acro{CBS}{complete basis set}
\acro{CC}{Coupled Cluster}
\acro{CMO}{Canonical Molecular Orbital}
\acro{CTCC}{Centre for Theoretical and Computational Chemistry}
\acro{CoE}{Centre of Excellence}
\acro{DC}{dielectric continuum}  
\acro{DFT}{Density Functional Theory}  
\acro{DIIS}{Direct Inversion of the Iterative Subspace}  
\acro{DKH}{Douglas-Kroll-Hess}
\acro{ESC}{Elimination of Small Components}
\acro{EFP}{effective fragment potential}
\acro{ECP}{effective core potential}
\acro{EU}{European Union}
\acro{FW}{Foldy–Wouthuysen}
\acro{GGA}{generalized gradient approximation}
\acro{GPE}{Generalized Poisson Equation}
\acro{GTO}{Gaussian-type orbital}
\acro{HF}{Hartree--Fock}  
\acro{HPC}{high-performance computing}
\acro{Hylleraas}[HC]{Hylleraas Centre for Quantum Molecular Sciences}
\acro{IEF}{Integral Equation Formalism}
\acro{IGLO}{individual gauge for localized orbitals}
\acro{KAIN}{Krylov Accelerated Inexact Newton}
\acro{KB}{kinetic balance}
\acro{KS}{Kohn-Sham}
\acro{LAO}{London atomic orbital}
\acro{LAPW}{linearized augmented plane wave}
\acro{LCAO}{Linear Combination of Atomic Orbital}
\acro{LDA}{local density approximation}
\acro{LMO}{Localized Molecular Orbital}
\acro{MAD}{mean absolute deviation}
\acro{maxAD}{maximum absolute deviation}
\acro{MM}{molecular mechanics}  
\acro{MCSCF}{multiconfiguration self consistent field}
\acro{MPA}{multiphoton absorption}
\acro{MPI}{Message Passing Interface}
\acro{MRA}{Multiresolution Analysis}
\acro{MW}{multiwavelet}
\acro{NAO}{numerical atomic orbital}
\acro{NeIC}{nordic e-infrastructure collaboration}
\acro{NMR}{nuclear magnetic resonance}
\acro{NP}{nanoparticle}  
\acro{OLED}{organic light emitting diode}
\acro{PAW}{projector augmented wave}
\acro{PBC}{Periodic Boundary Condition}
\acro{PCM}{polarizable continuum model}
\acro{PW}{plane wave}
\acro{QC}{quantum chemistry}  
\acro{QM/MM}{quantum mechanics/molecular mechanics}  
\acro{QM}{quantum mechanics}
\acro{RA}{Regular Approximation}
\acro{RCN}{Research Council of Norway}
\acro{RMSD}{root mean square deviation}
\acro{RKB}{restricted kinetic balance}
\acro{SAPT}{symmetry-adapted perturbation theory}
\acro{SC}{semiconductor}
\acro{SCF}{Self-Consistent Field}
\acro{SERS}{surface-enhanced raman scattering}
\acro{SI}{Supporting Information}
\acro{STSM}{short-term scientific mission}
\acro{WPREL}[WP1]{Work Package 1}
\acro{WPROP}[WP2]{Work Package 2}
\acro{WPAPP}[WP3]{Work Package 3}
\acro{WP}{Work Package}  
\acro{X2C}{exact two-component}
\acro{ZORA}{zero-order regular approximation}
\acro{ae}{almost everywhere}
\acro{BVP}{boundary value problem}
\acro{PDE}{partial differential equation}
\acro{RDM}{1-body reduced density matrix}
\acro{KAIN}{Krylov-accelerated inexact Newton}
\acro{FMM}{fast multipole method}
\acro{NUMA}{non-uniform memory access}
\end{acronym}

\makeatletter
\renewcommand{\@cite}[1]{
{$\!\! ^{(#1)}$}}
\makeatother

\newcommand{\orbital}{\ensuremath{\varphi}}

\newcommand{\rvec}{\ensuremath{\boldsymbol{r}}}

\newcommand{\scaling}{\ensuremath{\phi}}
\newcommand{\wavelet}{\ensuremath{\psi}}

\newcommand{\major}{\cellcolor{blue!100}}
\newcommand{\minor}{\cellcolor{blue!25}}

\title{The MRChem multiresolution analysis code for molecular electronic structure calculations: performance and scaling properties}

\author{Peter Wind}
\affiliation{Department of Chemistry, UiT - The Arctic University of Norway, N-9037 Tromsø, Norway}
\email{peter.wind@uit.no}

\author{Magnar Bjørgve}
\affiliation{Department of Chemistry, UiT - The Arctic University of Norway, N-9037 Tromsø, Norway}

\author{Anders Brakestad}
\affiliation{Department of Chemistry, UiT - The Arctic University of Norway, N-9037 Tromsø, Norway}

\author{Gabriel A. Gerez S.}
\affiliation{Department of Chemistry, UiT - The Arctic University of Norway, N-9037 Tromsø, Norway}

\author{Stig Rune Jensen}
\affiliation{Department of Chemistry, UiT - The Arctic University of Norway, N-9037 Tromsø, Norway}
\email{stig.r.jensen@uit.no}

\author{Roberto Di Remigio Eikås}
\affiliation{Algorithmiq Ltd, Kanavakatu 3C, FI-00160 Helsinki, Finland}

\author{Luca Frediani}
\affiliation{Department of Chemistry, UiT - The Arctic University of Norway, N-9037 Tromsø, Norway}
\email{luca.frediani@uit.no}

\begin{document}

\maketitle

\begin{abstract}
MRChem is a code for molecular electronic structure calculations, based on a multiwavelet adaptive basis representation. We provide a description of our implementation strategy and several benchmark calculations. Systems comprising more than a thousand orbitals are investigated at the \acl{HF} level of theory, with an emphasis on scaling properties. 
With our design, terms which formally scale quadratically with the system size, in effect have a better scaling because of the implicit screening introduced by the inherent adaptivity of the method: all operations are performed to the requested precision, which serves the dual purpose of minimizing the computational cost and controlling the final error precisely. Comparisons with traditional Gaussian-type orbitals based software, show that MRChem can be competitive with respect to performance.
\end{abstract}

\section{Introduction}

\acp{GTO} and more generally \acp{LCAO}\cite{Jensen.2013.10.1002/wcms.1123} are well established as a standard for \emph{ab initio} molecular electronic structure calculations. As their shape is closely related to the electronic structure of atoms, even very small basis sets with only a few functions per \ac{MO} can give reasonable results for describing molecular properties. However, for extended systems the description of each orbital still requires the contributions from the entire basis in order to ensure orthogonality. Without further precautions, even when using localized orbitals, a large proportion of the coefficients will be very small for those systems.

In a \ac{MRA} framework like \acp{MW}, the basis can adapt according to the function described (for an in-depth review of the \ac{MRA} method in the field of Quantum Chemistry, see Ref.~\citenum{Bischoff2019-mr}). The available basis is in principle infinite and complete, and, in practice, it is dynamically adapted to the local shape of the function and the required precision. 
This can require the real-space basis to comprise millions of elementary functions for each orbital. In this sense, the method starts with a big handicap compared to \ac{LCAO} basis sets. On the other hand, the exponential growth of available computational resources has in recent years enabled \ac{MRA} calculations on systems comprising thousands of atoms.\cite{Ratcliff2020-hk, Harrison2016-lo} 

Two main challenges need to be addressed in order to achieve adequate performance: the large number of operations to perform and the large memory footprint. The former is addressed by limiting the numerical operations to those that are strictly necessary to achieve the requested precision: rigorous error control at each elementary operation is the main strength of a \ac{MW} approach, enabling effective screening. The latter is achieved by algorithmic design: beyond a certain system size, not all data can be stored in local (i.e. fast access) memory. 
On modern computers, data access is generally more time consuming than the mathematical operations, especially if the data is not available locally.\footnote{The online interactive chart at \url{https://colin-scott.github.io/personal_website/research/interactive_latency.html} (Accessed 2022-10-29) is useful to understand the relative performance of on-chip \emph{vs.} off-chip memory accesses. In particular, it shows how the evolution of memory chips has been lagging behind that of CPUs.} 
The computer implementation must be able to copy large amounts of data efficiently between compute nodes and the algorithm must be devised to reuse the data already available on the compute node or in cache memory when possible. In this article, we will present the main implementation features of our \ac{MRA} code, MRChem \cite{mrchem}, to tackle those challenges, thus enabling calculations on large systems at the \ac{HF} level.

Beyond the effective thresholding that screens numerically negligible operations, the large computational cost is addressed by parallelization either at the orbital level or through real space partitioning. This dual strategy allows to minimize the relative cost of communication and the overall computational costs. Further, the most intensive mathematical operations are expressed as matrix multiplications, which allows for optimal efficiency.  
Using localized orbitals, the adaptivity of the \ac{MW} description will significantly reduce the computational effort required to compute the terms involving remote orbitals. 
Within a \ac{MRA} framework, the operators will exhibit an intrinsic sparsity, even if the orbitals have contributions from remote regions. This is achieved because the different length scales are naturally separated through the adaptive grid representation.
This opens the way for a method that scales linearly with system size ($N$), where the linearity arises naturally from the methodology, rather than being obtained with \emph{ad hoc} techniques, such as fast-multipole methods \cite{GREENGARD1987325} for Fock matrix construction, or purification of the density matrix. \cite{Helgaker2000-yb}

The large memory requirement is addressed by storing the data in a distributed ``memory bank", where orbitals can be sent and received independently by any CPU. The total size of the memory bank is then only limited by the overall memory available on the entire computer cluster.
Benchmark calculations at the \ac{HF} level show that MRChem is able to describe systems with thousands of electrons. The implementation exhibits near-linear scaling properties and it can also be competitive with state-of-the-art electronic structure software based on \ac{LCAO} methods.

\section{Solving the Hartree--Fock equations with multiwavelets}

We consider the \ac{SCF} equations of the \ac{HF} method:
\begin{equation}\label{eq:differential-SCF}
  \left(T + V \right) \orbital_i = \sum_{j}^{N_{\mathrm{occ}}} F_{ij} \orbital_j,
\end{equation}
where the Fock matrix elements $F_{ij}$ are obtained as
\begin{equation}\label{eq:fock-matrix-definition}
  F_{ij} = \Braket{ {\orbital_i}|{T+V}|{\orbital_j}}.
\end{equation}
In the above equations, $T = -\frac{1}{2}\nabla^2$ is the kinetic
energy operator, and the potential $V$ contains the nuclear-electron
attraction $V_{\mathrm{nuc}}$, the Coulomb repulsion $J$, and the exact exchange $K$. 
To solve the \ac{SCF} equations within a \ac{MW} framework, we rewrite the differential equation \eqref{eq:differential-SCF} as an integral equation:
\begin{equation}\label{eq:integral-SCF}
\orbital_i = -2 H^{\mu_i}\star\left[V\orbital_i - \sum_{j \ne i}^{N_{\mathrm{occ}}} F_{ij} \orbital_j\right],
\end{equation}
where $\star$ stands for the convolution product.
This form is obtained by recalling that the shifted kinetic energy operator $T-\epsilon$ has a closed-form Green's function:\cite{Kalos1962-ok,Beylkin2005-kg}
\begin{equation}\label{eq:helmholtz-green}
    (T-\epsilon)^{-1} = 2 H^{\mu}, \quad H^{\mu} = \frac{e^{-\mu r}}{r}, \quad \mu = \sqrt{-2\epsilon},
\end{equation}
By setting $\epsilon_i = F_{ii}$ and $\mu_i = \sqrt{-2F_{ii}}$, the diagonal term is excluded from the summation in Eq.~\eqref{eq:integral-SCF}.
It can be shown that iterating on the integral equation corresponds to a preconditioned steepest descent. Practical applications show that this approach is comparable in efficiency (as measured in the rate of convergence of the \ac{SCF} iterations) to more traditional methods.\cite{Harrison2004-fs}

A \ac{MW} representation does not have a fixed basis. It is therefore not possible to express functions and operators as vectors and matrices of a predefined size, and the virtual space is to be regarded as infinite. Still, each function has a finite, truncated representation defined by the chosen precision threshold, but this representation will in general be different for different functions. It is therefore necessary to develop working equations which allow the direct optimization of orbitals. On the other hand, only occupied orbitals are constructed and the coupling through the Fock matrix in Eqs. \eqref{eq:differential-SCF} and \eqref{eq:integral-SCF} is therefore limited in size. Another advantage is that, to within the requested precision, the result can be formally considered exact, and exploiting formal results valid in a complete basis -- most notably the Hellmann--Feynman theorem -- becomes straightforward.\cite{Yanai2004-tw}

Differential operators such as the Laplacian pose a fundamental problem for function representations which are not continuous, and a naive implementation leads to numerical instabilities. Robust approximations are nowadays available\cite{Anderson2019-bx}, but for repeated iterations avoiding the use of the Laplacian operator is still an advantage. This is the main reason for using the integral form in Eq.~\eqref{eq:helmholtz-green} rather than the differential form.

\section{Implementation details and parallelization strategy}

\ac{MW} calculations can be computationally demanding. Both the total amount of elementary operations, the large amount of memory required and the capacity of the network are important bottlenecks. 

In practice, a supercomputer is required for calculations on large systems.
For example the representation of one orbital can demand between 10 and 500MB of data, depending on the kind of orbital and the precision requested (see Section \ref{sizes} for details). Moreover, several sets of functions are necessary in a single \ac{SCF} iteration (orbitals at the previous iterations, right-hand side of equations etc.). Large systems will eventually require more memory than is locally available on a single compute node in a cluster, and distributed data storage becomes mandatory. At the same time, the \ac{SCF} equations will require pairs of orbitals, i.e. all the data must at some point be accessible locally.

For an efficient solution scheme on a cluster architecture, the problem needs to be partitioned into smaller portions that can be computed with as little transfer of data as possible between the different compute nodes of the cluster.

\subsection{Localized versus canonical orbitals}

Although localized and canonical orbitals represent the same $N$-electron wave function, the numerical operations required to solve the equations will have different scaling properties in a canonical or localized representation.
Because of the inherent adaptivity of the \ac{MW} approach, multiplying two orbitals, or applying an operator onto an orbital can be done more efficiently if the orbitals are localized in space. Even if the orbitals have non-negligible values in a large portion of space, the adaptivity properties will make sure that the calculations in the parts where the orbitals are small are done at a coarser level of accuracy, and therefore faster than in the regions with larger values.

We use the Foster--Boys localization criterion.\cite{Foster1960} The unitary, localizing orbital rotation is computed by minimizing the sum of the occupied orbitals' second central moment.\cite{Hoyvik2016-cw} See also sections \ref{sec:localvsorb} and \ref{sec:N3terms} for more details on the practical implementation.

\subsection{Local (real-space) versus orbital representation}
\label{sec:localvsorb}

\begin{figure}[htb]
\centering
\includegraphics[width=.6\textwidth]{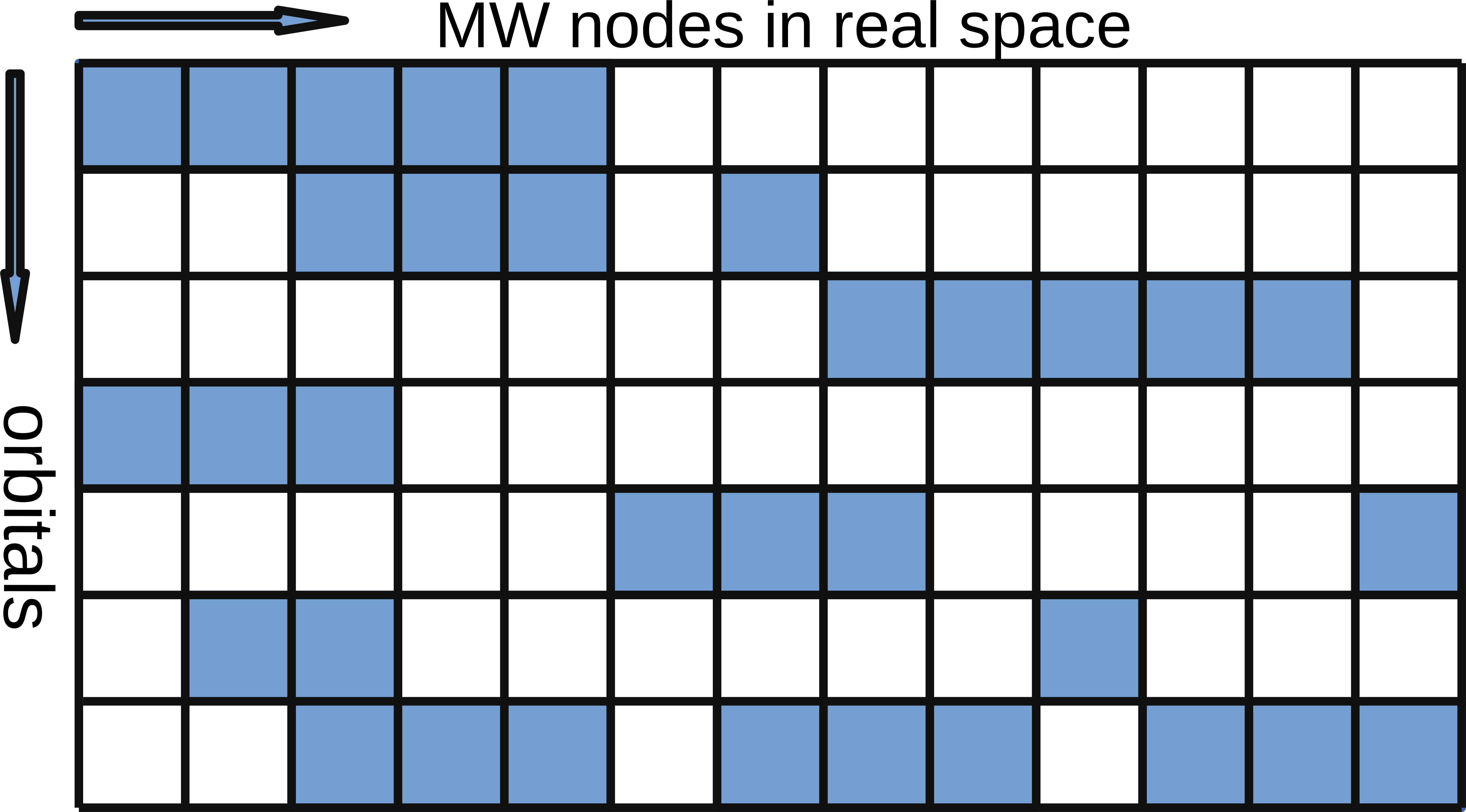}
\caption{Real-space \emph{vs.} orbital representation. A dark cell represent a \ac{MW} node occupied by the corresponding orbital. In an orbital representation (rows) the data from each row is stored and retrieved together. In a local representation (columns), the data from each column is stored and used together. The most convenient representation depends on the operation at hand and it is therefore best to be able to switch from one to the other.}
\label{fig01}
\end{figure}

Two main strategies can be adopted to distribute the memory storage requirements for orbitals throughout the cluster. The first strategy follows \emph{verbatim} the mathematical expressions: the \ac{HF} equations are written in terms of orbitals, and therefore it is natural to keep the data representing each orbital as a separate C++ structure. The second strategy, is based on a real-space partitioning: contributions from all the orbitals in a given portion of real space are stored together, i.e. as a separate C++ structure that is computationally cheap to retrieve (see Figure~\ref{fig01}). 
Both strategies have advantages and disadvantages, and in practice it is convenient to switch between them depending on the given operation.

Taking as an example the computation of the Fock matrix elements in Eq.~\eqref{eq:fock-matrix-definition},  we will show why it is advantageous to shift to a real-space partitioning approach.\footnote{A similar development has happened for the \ac{LCAO} methods: they first were based on \acp{MO}, but when the focus shifted towards large systems, it was soon realized that efficient implementations demanded a design based on using \acp{AO} directly.\cite{Helgaker2000-yb}}
Operator matrix elements are evaluated as 

\begin{equation}
  V_{ij} = \braket{\orbital_{i} | {\hat V} |\orbital_j}  = \braket{\orbital_i | {\hat V} \orbital_j} 
\end{equation}
where $\orbital_i$ and $\orbital_j$ are occupied orbitals and ${\hat V}$ is any one-electron operator, such as the nuclear potential operator. If the application of the operator ($\ket{{\hat V} \orbital_j},\,\,\forall j$) is considered as a $O(N_{\mathrm{occ}})$
operation, computing the entire matrix scales formally as $O(N_{\mathrm{occ}}^2)$. 

Both the $\orbital_i$
and the result of the operator application $\tau_j = \hat{V} \orbital_j$
are represented either through the \emph{compressed} or \emph{reconstructed} representation, as described in the Section 1 of the Supporting Information. For instance the compressed representations for $\orbital_i$ and $\tau_i$ would yield respectively
\begin{equation}\label{eq:compressed_orb}
  \orbital_i = \sum_{p} c^i_p \scaling_p + \sum_{np} d^i_{np} \wavelet_{np} \quad\text{and}\quad \tau_j = {\hat V} \orbital_j = \sum_{p} t^j_p \scaling_p + \sum_{np} w^j_{np} \wavelet_{np},
\end{equation}
where $n$ 
runs over all the included \ac{MW} nodes (see \ac{SI} for details about representing functions using \ac{MW}).

When representing functions in 3 dimensions,
using a tensor-product basis and localized orbitals, both $p$ and the number of \ac{MW} nodes, are typically of size $\sim 10^4$. 
To perform a scalar product, it is sufficient to recall that the scaling/wavelet basis is fully orthonormal:
\begin{equation}\label{eq:wavelet_othrogonality}
  \braket{\scaling_{p} | \scaling_{p'}}  = \delta_{pp'}, \quad   \braket{\wavelet_{np} | \wavelet_{n'p'}}  = \delta_{nn'}\delta_{pp'}, \quad   \braket{\scaling_{p} | \wavelet_{n'p'}}  = 0, 
\end{equation}
which then yields the following result for an operator matrix element between two orbitals:
\begin{equation}
  \braket{\orbital_i|{\hat V} \orbital_j} = \sum_{p} c^i_{p} t^j_{p} + \sum_{np} d^i_{np} w^j_{np}
\end{equation}
where now the sum over the \ac{MW} nodes $n$ only extends to those which are present for both functions.

Ignoring for simplicity the scaling contribution which has negligible impact on the performance, this operation is equivalent to a matrix multiplication $D \cdot W^T$ with $D_{ia} = d^i_{np}$ and $W_{ja} = w^j_{np}$, and $a$ is the compound index $a=np$. However, for large systems it is not possible to hold the full matrices in local memory.
The \ac{MO}-like approach is then to sum over all compound indices $a$ for each pair $i,j$ independently. In this way each pair of coefficients $d^i_{np}$ and $w^k_{np}$ is used only once after each memory fetch. On modern computers, memory access is much more time-consuming than the mathematical operations themselves, up to a factor 100 for most CPU-based systems. A faster implementation is to consider a specific \ac{MW} node $n$ and perform the product for all $i$, $j$ and $p$ values within this node:
\begin{equation}\label{equ:sum}
  \braket{\orbital_i | {\hat V} |\orbital_j}_{n} 
  = \sum_{p} \left(d^i_{p} w^j_{p}\right)_{n}, \quad\forall i,j,
\end{equation}
this results in a series of independent matrix multiplications, thus fully utilizing the capabilities of the computer. 

To perform the operation in this way, the storage of the orbitals in memory must be adapted. Instead of storing each orbital separately (orbital or MO representation), the values of all the orbitals at specific positions in space (\emph{i.e.}~a \ac{MW} node) must be easily accessible (local representation: all the data covering a specific \ac{MW} node is stored on the same compute node, as an independent C++ structure). Switching from an orbital to a local representation is a $O(N)$ operation, where $N$ measures the system size, typically the number of orbitals.
If all $i$ and $j$ are included, the number of operations still scales formally as the square of the number of orbitals. However for a given \ac{MW} node $n$, the products only needs to be explicitly performed if both functions (indices $i$ and $j$) have a non-negligible contribution to the \ac{MW} node $n$. When orbitals are localized, the number of such contributions will eventually scale linearly with the system size.

Linear combinations of orbitals can be addressed similarly. Let us consider the linear combination
\begin{equation}
  \tilde{\orbital}_i = \sum_{j} A_{ij} \orbital_j
\end{equation}
where $A_{ij}$ is the coefficient matrix of the transformation.
In a linear transformation, each \ac{MW} node $n$ can be transformed independently, to obtain the corresponding coefficients for $\tilde{\orbital}_i$:
\begin{equation}
  w^i_{np} = \sum_{j} A_{ij} d^j_{np}
\end{equation}
Here also, the number of localized orbitals contributing to a specific \ac{MW} node $n$ grows as $O(1)$ with system size. 
When constructing $\tilde{\orbital}_i$, the grid is successively refined, so that only values corresponding to \ac{MW} nodes with wavelet norms above the required threshold are actually computed.
Moreover, due to linearity this operation is carried out both for scaling and wavelet coefficients, thus avoiding the need to perform any \ac{MW} transforms.

\subsection{Adaptivity and precision thresholding}\label{sec:adaptivity-screening}

The most powerful idea of \ac{MRA} is to limit the precision of intermediate operations to what is strictly necessary to achieve the requested precision in the final result. This guiding principle can be exploited at several levels:

\begin{enumerate}
    \item Elementary mathematical operations on a function in a multiwavelet representation will automatically refine the local basis (\emph{i.e.}~grid) of the resulting function, but only until the required precision is achieved. The computational effort required will vary according to the position and shape of the functions or operators. For example, the product of two functions which are localized in different parts of space, will normally be done faster than the square of each of them.
    \item Only those parts of functions with absolute value (norm of the \ac{MW} node) above a defined threshold will be retrieved before even processing them; this is used \emph{e.g.}~in the scheme presented in section \ref{sec:localvsorb}.
    \item A \ac{SCF} calculation can be initiated at low precision and the final result can be used as starting guess for a calculation with higher precision. In contrast to finite \ac{AO} basis sets, which are usually not subsets of each other, a multiwavelet function computed at a given precision, can be directly used as a valid representation at a different precision.
\end{enumerate}

\subsection{Exchange screening}
\label{Xscreen}

The exchange terms require the evaluation of

\begin{equation}
\label{equ:ex}
  K_i \orbital_j = \orbital_i(\rvec) 
  \left[
  \int \frac{\orbital^*_i(\rvec')\orbital_j(\rvec')}{|\rvec-\rvec'|} \, \mathrm{d}\rvec'
  \right]
\end{equation}
 for all $\frac{N_{\mathrm{occ}}(N_{\mathrm{occ}}-1)}{2}$ orbital pairs $i,j$.

The evaluation of equation (\ref{equ:ex}) can be divided into 3 operations:
\begin{itemize}
\item Product of $\orbital^*_i(\rvec')$ and $\orbital_j(\rvec')$
\item Application of the Poisson operator, $\int \frac{\orbital^*_i(\rvec')\orbital_j(\rvec')}{|\rvec-\rvec'|} \, \mathrm{d}\rvec'$
\item Product with $\orbital_i(\rvec)$, $\orbital_i(\rvec) \left[\int \frac{\orbital^*_i(\rvec')\orbital_j(\rvec')}{|\rvec-\rvec'|} \, \mathrm{d}\rvec'\right]$
\end{itemize}

The application of the Poisson operator gives a long range potential that decays as $\frac{1}{r}$. However, if the orbital $\orbital_i(\rvec)$ is localized, only the part of the potential where $\orbital_i(\rvec)$ is large enough needs to be computed. More precisely, the potential needs only to be computed up to a precision which is inversely proportional to the value of $\orbital_i(\rvec)$ at that point. The Poisson operator can therefore use a locally defined threshold, where the required precision of the operator application is determined using local information from a \emph{proxy} for the result function.

As for all \ac{MW} representations of functions, the output of a Poisson operator application uses an adaptive grid: when a new \ac{MW} node is computed, the norm of the wavelet part is compared to the precision threshold to decide whether to further refine the representation or not. In general, the precision threshold is fixed: it is one and the same over the entire space. For the \ac{HF} exchange, we already know that the result of convolution with the Poisson kernel will only be used as part of the product with $\orbital_i(\rvec)$. Thus, the precision threshold can be multiplied by the local value of the norm of $\orbital_i(\rvec)$ (the reference function). This procedure will yield a grid which is adapted to the final result, with increased precision corresponding to the regions with non-negligible values of $\orbital_i(\rvec)$.

Several reference functions can be provided, and the precision at a given \ac{MW} node is determined by the largest value of the norm of those functions at the node. This allows to treat the two terms $\orbital_i(\rvec) \left[\int \frac{\orbital^*_i(\rvec')\orbital_j(\rvec')}{|\rvec-\rvec'|} \, \mathrm{d}\rvec'\right]$ and  $\orbital_j(\rvec) \left[\int \frac{\orbital^*_i(\rvec')\orbital_j(\rvec')}{|\rvec-\rvec'|} \, \mathrm{d}\rvec'\right]$ using the same potential $\int \frac{\orbital^*_i(\rvec')\orbital_j(\rvec')}{|\rvec-\rvec'|} \, \mathrm{d}\rvec'$.  Also complex functions can take advantage of this feature, since the real and imaginary parts of $\orbital_i(\rvec)$ can then use the same potential.

Without screening, the number of exchange terms grows as $N^2_{\mathrm{occ}}$ and each term requires the application of the Poisson operator. Distant orbitals that do not overlap, do not contribute, and therefore proper handling of the screening will be reflected in a cost of the exchange terms that grows linearly in size for large enough systems (if the required overall absolute precision is held constant). 
Additionally, if the product of $\orbital^*_i(\rvec')$ and $\orbital_j(\rvec')$ is small enough, the calculation of the corresponding term can be avoided altogether. As shown in Section \ref{sec:screen} this leads to an effective long range screening.

\subsection{$O(N_{\mathrm{occ}}^3)$ terms}
\label{sec:N3terms}

In our implementation there are two steps with an operation count that formally scales with the third power of the number of (occupied) orbitals: the localization
and the orthogonalization steps. The former involves the product of matrices with sizes $N_{\mathrm{occ}} \times N_{\mathrm{occ}}$; the latter performs the diagonalization of a Hermitian matrix and subsequent matrix multiplications, both with cubic formal scaling.

Matrix operations can be performed extremely efficiently on any computer using standard linear algebra libraries, and those terms will not contribute significantly to the total computation time before considering sizes larger than roughly $N_{\mathrm{occ}}=1000$ for serial matrix multiplications, or $10^4$-$10^5$ for parallel, distributed matrix multiplications (not yet implemented). We have therefore not yet rendered such operations faster than their formal scaling.

We note that both steps are not specific to the multiwavelet approach. As a matter of fact, \ac{LCAO} approaches may require those steps to be performed on the virtual orbitals too. For large systems, it might also be possible to exploit the numerical sparsity of the computed matrices.


\subsection{Parallelization}

MRChem is parallelized using the \ac{MPI} standard\cite{Pacheco1997-qq,Gropp2014-dz,Gropp2014-qf} and OpenMP threading.\cite{Mattson2019-gl,Van_der_Pas2017-wq}
The MPI processes are partitioned into three groups which will be described in the following section.

\subsubsection{MPI ``worker" processes} 
They perform the mathematical operations. These processes have also access to multiple threads for OpenMP parallelization. The large majority of the cores are assigned to worker processes.

The low-level operations (on one or two orbitals or \ac{MW} functions, and on \ac{MW} nodes) are included  in the MRCPP library. MRCPP will perform tree operations as a set of operations on \ac{MW} nodes. Those operations will then be performed in parallel by OpenMP threads. The thread parallelism is transparent for the library user, who will only need to set the number of OpenMP threads.

\subsubsection{Memory bank processes} 
These processes are only receiving and distributing data to the workers, they do not perform major mathematical operations and are not threaded. Bank processes will impose a high load on the network; it is therefore important to distribute them on all compute nodes in order to avoid saturation of the local network. Since the memory bank processes are different from the worker processes, the worker processes can access all the data without direct communication with other workers (in effect, one-sided memory access). Efficient transfer of data is the key to a successful parallelization scheme.

In the \ac{MW} approach, orbitals are represented as tree structures. The root nodes of the \ac{MW} tree cover the entire space of the system, and each subdivision defines branches. To facilitate the transfer of those functions, the tree structure and the \ac{MW} node data are stored in piecewise contiguous memory. This allows the data to be sent efficiently among MPI processes: only the position in memory of the start and end of the pieces are required, alongside some metadata. A \ac{MW} tree can then be transferred without intermediate buffering and with little overhead. 
The MRCPP library is able to transfer \ac{MW} trees (or orbitals) between \ac{MPI} processes easily and efficiently.

\subsubsection{Task manager} 
In a localized orbital approach, given the matrix representation $\mathbf{V}$ of an operator $\hat{V}$, many of its matrix elements $V_{ij} = \braket{\orbital_i|{\hat V} \orbital_j}$ will be small. The \ac{MRA} framework implemented in the MRCPP library is inherently adaptive and small elements are computed faster. To exploit such a feature, it is important to equip the library with a dynamical distribution of tasks.
In practice, the matrix $\mathbf{V}$ is partitioned into blocks, each block representing one task to be performed by a ``worker". In a local representation, each \ac{MW} node is simply assigned to a separate task. When a ``worker" is done with one task it can ask the~\emph{task manager} for a new task until all tasks are completed.

One single MPI process is reserved for managing the task queue. The only purpose of this process is to keep account of the tasks that have been assigned and the tasks left, and distribute unsolved tasks to the workers. It does only send and receive small amounts of data at a time (a few bytes) and will not be saturated with requests.

\section{Benchmark results}
In this section we will show how MRChem performs for some simple systems of different size. MRChem is still under development and the results shown here are just a snapshot of the present situation. The goal is not to determine the limitations of the model, but rather to show that it behaves as intended and that the development is going in the right direction.

All benchmark runs presented here were performed on ``Betzy", a Bull Sequana XH2000 computer cluster, with 128 cores and 256 GiB memory per compute node, and InfiniBand HDR 100 interconnect.\cite{Betzy}

All timings given in this section are subject to a degree of uncertainty. There are multiple reasons for this. The runs have been executed in a MPI parallel framework on a supercomputer shared with other users. Several algorithms exploit a dynamic allocation of tasks; as a consequence the same calculations can have a different workload distributions in two repeated identical runs. The network capacity may also be affected by other users. Finally, the time spent in different parts of the code may not be well defined: in a parallel framework the same code will start and end at different times for different processes. In principle the time for one block should include the time for all processes to be finished, however in practice processes are allowed to continue without waiting, until they need data which is not yet available.
The time for a similar \ac{SCF} cycle may vary by up to 10\%. However, this accuracy should be sufficient and reflect the main trends faithfully. 

\subsection{Scaling with system size}

\begin{figure}
\centering
\includegraphics[width=\textwidth]{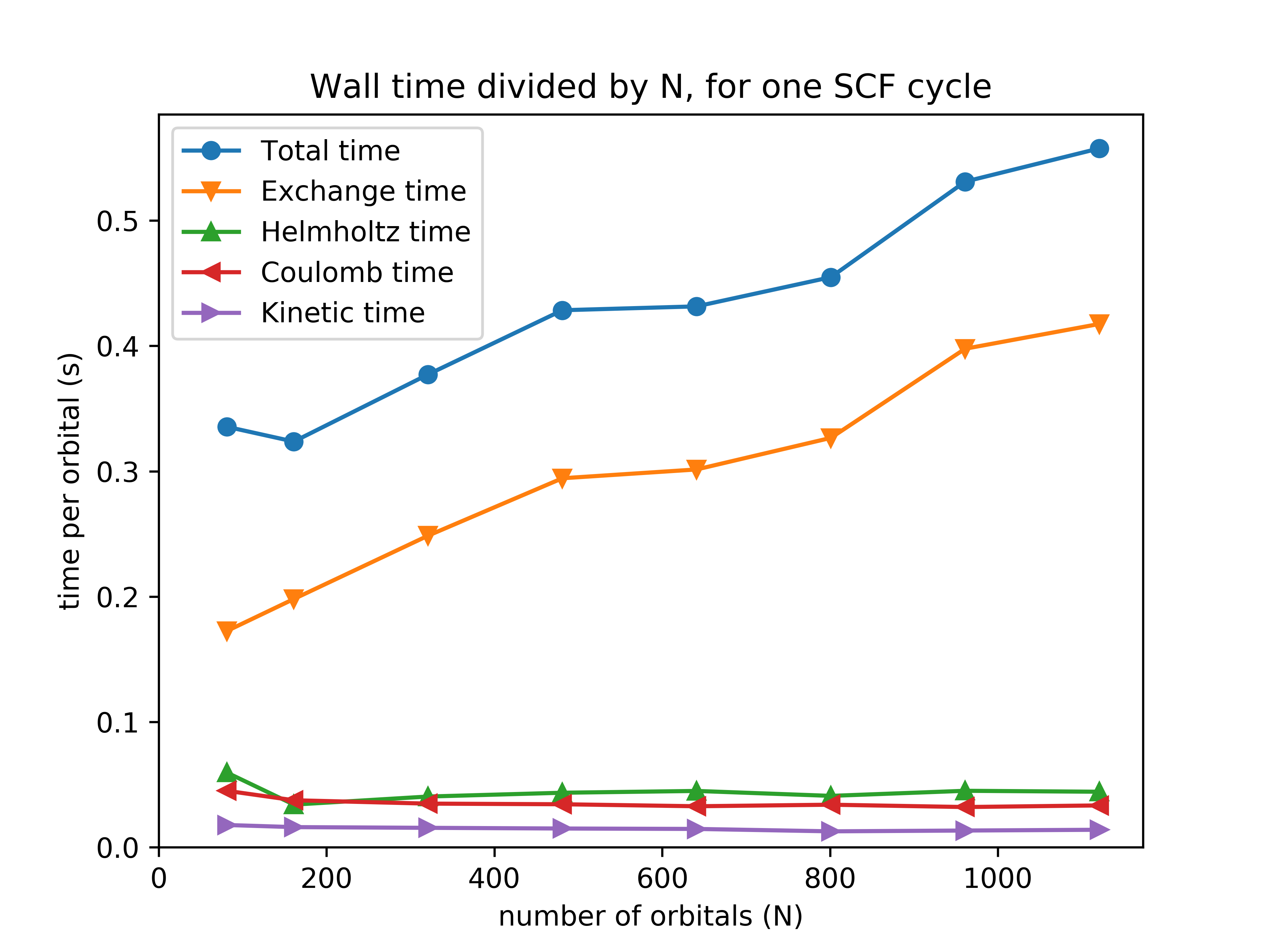}
\caption{Timings divided by number of orbitals for one \ac{SCF} cycle for the linear alkanes \ce{C_{n}H_{2n+2}} ($n=20, 40, 80, 120, 160, 200, 240, 280$). All calculations performed using 8 compute nodes. \emph{Total time} refers to the time spent in one Hartree-Fock \ac{SCF} cycle, when it is close to convergence. \emph{Exchange time} refers to the time spent for computing the Exchange contributions. \emph{Coulomb time} refers to the time used for computing the Coulomb potential of each orbital, summing them up and distributing the total potential to all the MPI processes. \emph{Kinetic time} refers to the time spent for computing the kinetic energy of each orbital and \emph{Helmholtz time} the time for constructing and applying the Helmholtz operator. Constant time indicates linear scaling.}
\label{fig02}
\end{figure}

Figure \ref{fig02} shows computation times divided by the number of orbitals for alkanes of different lengths. For a linear scaling method, the time per orbital should be approximately constant.  The time per orbital less than doubles going from the smallest (81 orbitals) to the largest (1121 orbitals) alkane. This shows that the scaling with system size is mostly linear, with a small non-linear contribution.  


The application of the Exchange operator is by far the most time-consuming part of the \ac{SCF} process. The other parts are showing close to linear scaling behavior.
The number of non-negligible exchange contributions should ideally grow linearly with system size for large enough systems, and therefore the exchange computation time per orbital should be constant. However, exchange contributions are the result of a sum of terms, and if the number of terms increases, the accuracy of each individual term must be increased in order to achieve a fixed absolute precision in the final result. In our implementation, this is taken care of by increasing the required precision of the individual terms by a factor proportional to the square root of the number of orbitals. This is the main reason why our run times are exhibiting a slight non-linearity.

\subsection{Scaling with precision}
 
In Table \ref{tab:prec} we have performed calculations with different precisions on two molecules, valinomycine (\ce{C54H90N6O18}, 300 orbitals) and the gramicidin dimer (\ce{C198H276N40O34}, 1008 orbitals). MW5 $(10^{-5})$, MW6 $(10^{-6})$, and MW7 $(10^{-7})$ denote the value of the user-defined precision parameter. 
Our examples show a factor of 2.5 increase in computing time for each ten-fold increase in precision (one additional digit). A similar test with \acp{GTO}, performed increasing the basis set size, shows a much less favourable scaling: see examples from Table \ref{tab:MRC_ORCA_LSD} below.

\begin{table}[htb]
    \caption{Time (in seconds) for different terms in the \ac{SCF} cycle. 16 and 64 compute nodes used for valinomycine (\ce{C54H90N6O18}, 300 orbitals) and gramicidin dimer (\ce{C198H276N40O34}, 1008 orbitals), respectively. Total timing is dominated by the \ac{HF} exchange. The additional cost for each precision increase is roughly a factor 2.5-3 for all contributions.
    }
    \label{tab:prec}
    \begin{tabular}{c c S[table-format=3.1] S[table-format=3.1] S[table-format=2.1] S[table-format=2.1] S[table-format=4.1]}
    \toprule
    & & {Exchange} & {Coulomb} & {Kinetic energy} & {Helmholtz} & {Total} \\
    \midrule
 \multirow{3}{*}{Valinomycine} & MW5 &   71.4 &  9.2 &  2.7 &  7.4 &  97.4 \\
                               & MW6 &  182.4 & 25.3 &  6.7 & 21.3 & 251.2 \\
                               & MW7 &  442.8 & 78.8 & 17.1 & 66.4 & 644.4  \\
    \midrule

 \multirow{3}{*}{Gramicidin} & MW5 & 115.2 &  18.4 &  4.2 &  8.0 &  163.7 \\  
                             & MW6 & 312.0 &  45.8 & 14.6 & 25.5 &  428.9 \\ 
                             & MW7 & 813.6 & 127.8 & 28.5 & 83.1 & 1196.9 \\ 
  \bottomrule
    \end{tabular}
\end{table}

\subsubsection{Size of orbital representation}

\label{sizes}

In a localized approach, the orbital functions have negligible values in remote regions of space. In a \ac{LCAO} approach (if no special filtering is applied) the representation of an orbital will span the entire represented space, since the basis will increase with the system size. In the \ac{MRA} approach orbitals will essentially be represented using locally defined functions and their size will depend only weakly on the size of the entire system.
This is confirmed directly in our calculations, see Table \ref{tab:orbsizes}. Those results were obtained using a threshold parameter of $10^{-5}$ (MW5), $10^{-6}$ (MW6) and $10^{-7}$ (MW7). 
The individual size of the orbitals will of course vary greatly according to their type; for example core orbitals are simpler and therefore require less data at a given precision, more diffuse function have more features and their representation requires more data. In the molecules presented here, the largest orbital may be two to four times larger than the average, and the smallest a third of the average size.
Our empirical observations show a factor in the range  2 - 2.2 increase in orbital size for each ten-fold increase in precision. 
An increase in the number of coefficients describing the orbitals will clearly increase the computational cost of each operation using this orbital. In addition an increase in precision will also extend the range of orbitals with non-negligible interactions, \emph{i.e.} demands less aggressive screening. 
The number of orbitals combined with the average size of the orbitals, can be taken as an indicator of the amount of computational resources (memory, CPU time, and network) required to solve the equations. 

\begin{table}[htb]
\caption{Average orbital size for different molecules in megabytes (MB) for increasingly tighter precision parameters. The average orbital size is evaluated in in terms of the number of \ac{MW} nodes or coefficients required. For the largest molecules in the alkane series is almost independent of the molecule size.} 
\label{tab:orbsizes}
\begin{tabular}{cccccccccc} 
\toprule
& \ce{C20H42} & \ce{C40H82} & \ce{C80H162} & \ce{C160H322} &  valinomycine & gramicidin\\
\midrule
MW5 orbital size (MB)  &  35 &  43 &  43 &  46 &  49 &  55\\
MW6 orbital size (MB)  &  80 &  85 &  90 &  92 & 115 & 136\\
MW7 orbital size (MB)  & 158 & 167 & 176 & 180 & 245 & 306\\
\bottomrule
\end{tabular}
\end{table}

\subsection{Scaling with number of compute nodes}

\begin{figure}[htb]
\centering
\includegraphics[width=\textwidth]{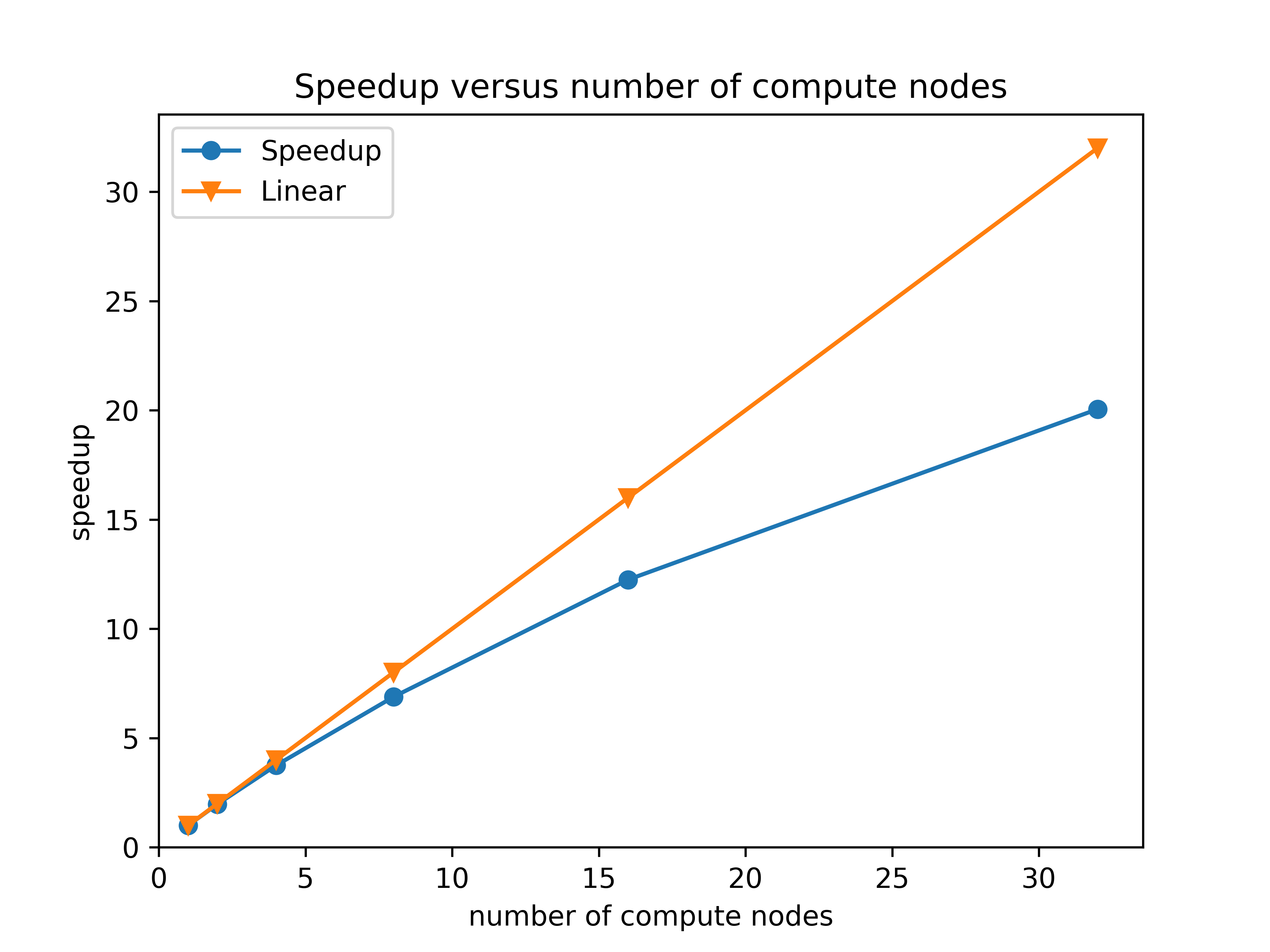}
\caption{The speedup is the inverse of the time for one Hartree-Fock cycle for the valinomycine molecule (\ce{C54H90N6O18}) at MW4 precision, relative to the time used on one compute node. Each compute node hosts 12 MPI processes, four of which are task manager/memory banks (single threaded) and eight are worker processes (15 threads each).}
\label{fig03}
\end{figure}

The time to perform a full calculation will normally decrease when using more computational resources (compute nodes). Ideally a doubling of the number of computational resources would half the calculation time. However a larger number of compute nodes implies that an increasing fraction of time must be spent distributing the data around. 
We define the speedup as the time for a calculation on one compute node divided by the time using $N_{\text{nodes}}$. Ideally the speedup is equal to the number of compute nodes (linear).    
In Figure \ref{fig03} we show the speedup for the valinomycine molecule at MW4 precision for $N_{\text{nodes}} = 1, 2, 4, 8, 16, 32$. 

For 32 compute nodes the time for one SCF cycle is 20 times faster than using 1 compute node, showing a relative efficiency of $20/32 \simeq 63$\%.
For larger systems or higher precision, the efficiency is expected to increase.

\subsection{Implicit screening} 
\label{sec:screen}

It is instructive to see more in detail how the adaptivity of \ac{MRA} leads to a significant reduction of the computational cost. We will consider linear alkanes of different sizes, as they are easier to compare.

As a first illustration we will consider the evaluation  of the Fock matrix, Eq.~\eqref{eq:fock-matrix-definition} using the formulation from Equation \eqref{equ:sum}:
\begin{equation}
\label{equ:sums}
 \braket{\orbital_i | {\hat F} |\orbital_j}_n = \sum_{p} c^i_{np} d^j_{np} \quad\forall i,j, 
\end{equation}
where $i$ and $j$ run over the occupied orbitals, $n$ runs over all \ac{MW} nodes ($n$ collects all values of $s$ and $l$ present in the tree) and
$p$ runs over the \ac{MW} basis within that node, i.e. 3-dimensional polynomials. In the calculations presented in this paper, polynomials up to 7th order in each direction are used, leading to a total of 4096 3-dimensional polynomials (wavelet and scaling) for each \ac{MW} node. We underline that this choice might be sub-optimal for high precision calculations (MW7) but we have decided to keep the parameter fixed for ease of comparison.

In theory, the \ac{MW} basis comprises all polynomials at the locally finest scale, in the union of all the orbital grids. This basis can be very large, but we never explicitly construct it in full. In Table \ref{tab:adapF} we show the total number of \ac{MW} nodes present in this basis (i.e. number of indices $n$). As expected, this number grows linearly with the size of the system.
 
\begin{table}[htb]
\caption{Number of terms and timings for the Fock matrix calculation 
for \ce{C_{n}H_{2n+2}}. Precision MW4. The number of fully computed terms increases faster than linearly, but the fraction of time to perform the corresponding multiplications is small. 
}
\label{tab:adapF}
\begin{tabular}{S[table-format=3] S[table-format=5] | S[table-format=5] S[table-format=2.1] S[table-format=2.2] | S[table-format=3] S[table-format=3]}
\toprule
{$n$} & {\ac{MW} nodes} & {blocks} &  {computed} & {neglected (\si{\percent})} & {fetch (\si{\milli\second})} & {multiply (\si{\milli\second})} \\
\midrule
 20  &  6472 &    42M &  1.6M & 96.33 & 164 &  19 \\ 
 40  & 12936 &   335M &  3.8M & 98.86 & 340 &  38 \\ 
 80  & 26120 &  2691M & 10.5M & 99.61 & 532 &  97 \\ 
160  & 52136 & 21422M & 30.8M & 99.85 & 981 & 270 \\ 
\bottomrule
    \end{tabular}
\end{table}
 
Let us now define a \emph{block} as one set of $\lbrace i,j,n\rbrace$ indices. The total number of blocks  is the number of \ac{MW} nodes $n$ multiplied by the square of the number of orbitals. For each block, Eq.~\eqref{equ:sum} defines the multiplication of the corresponding coefficients for all the polynomials defined within a \ac{MW} node. The total number of blocks increases with the third power of the system size. However, if a given \ac{MW} node is present for one orbital $i$, but not for orbital $j$, it will not contribute to the corresponding matrix element, and can be omitted. This follows directly from the orthogonality of the underlying wavelet basis, Eq.~\eqref{eq:wavelet_othrogonality}. For localized orbitals, $i$ and $j$ will contribute only to limited subsets of nodes and a large proportion of terms are either zero or have negligible numerical values in relation to the requested precision. They can therefore be discarded without being evaluated. This is a direct consequence of the disjoint support of the \ac{MW} basis, combined with localization.
 
As shown in Table \ref{tab:adapF}, the number of non-negligible terms is still increasing faster than linear. 
This is not necessarily synonym with computation time increasing faster than linear. To perform the calculations, the code must first fetch the relevant data in the bank, \emph{i.e.}~the matrices $c^i_{np}$ and  $d^j_{np}$ for given $n$. Then the two matrices are multiplied. As shown in the table the total time used to fetch the data grows slightly slower than linearly. 
For large systems, the total time to perform the matrix multiplication is proportional to the number of computed terms, and grows faster than linear. On the other hand, the matrix multiplication part is done so efficiently that it takes only a small fraction of the total time. In fact in our implementation we did not even use threaded matrix multiplication (yet); for large systems, an order of magnitude could be gained on the timings for the matrix multiplication part using threaded matrix multiplication. 

\begin{table}[htb]
\caption{Number of terms in the Exchange calculation for the \ce{C_{n}H_{2n+2}} series. Precision MW4. The number of terms fully computed becomes proportional to the number of orbitals.} 
\label{tab:adapEx}
\begin{tabular}{S[table-format=3] S[table-format=3] S[table-format=6] S[table-format=5] S[table-format=2]}
\toprule
{$n$} & {Number of orbitals} & {non-diagonal terms} & {fully computed} & {fraction neglected} \\
\midrule
 20 &  81 &   3240 &  1146 & 65\% \\ 
 40 & 161 &  12880 &  2548 & 80\% \\ 
 80 & 321 &  51360 &  5194 & 90\% \\ 
160 & 641 & 205120 & 10404 & 95\% \\ 
\bottomrule
\end{tabular}
\end{table}

Table \ref{tab:adapEx} shows the total number of exchange terms and how many of them are actually fully computed. Since the terms are computed in pairs and the number of non-diagonal pairs is $N(N-1)/2$, their number scales quadratically.    
The computation of the exchange terms starts with the product of two orbitals. If the product has a norm smaller than a threshold, it can be neglected without having to apply the Poisson operator (which is the computationally expensive part). We see that a large fraction of terms is effectively neglected, and the number of terms remaining scales linearly.

\subsection{Comparison with ORCA  and LSDalton  }

Table \ref{tab:MRC_ORCA_LSD} shows the computation time for one
\ac{SCF} cycle for the valinomycine molecule computed at the
\ac{HF} level. The accuracy of the total energy and atomization
energies are compared with the ORCA and LSDalton results using a pc-1,
pc-2 and pc-3 basis sets.\cite{Jensen2001-qo, Jensen2002-pp} Calculations with the pc-4 basis either did not run or did not converge.

\begin{table}[htb]
    \caption{Precision and performance comparison with ORCA and LSDalton, for one \ac{SCF} cycle for the valinomycine molecule (\ce{C54H90N6O18}). $E$ is the total energy and $\Delta E$ is the atomization energy. The error is defined as the difference with the MW7 results. Wall time per \ac{SCF} cycle is reported. $N_{\mathrm{P}}\,(N_{\mathrm{C}})$ is the number of GTO primitive (contracted) functions. All computation where done on 4 compute nodes (\emph{i.e.} 512 cores), except for MRChem MW5, MW6 and MW7 which were performed using 16 compute nodes (timings are marked with $^{*}$). Their timings have been multiplied by a factor 4 for ease of comparison.}\label{tab:MRC_ORCA_LSD}
    \begin{tabular}{llcccccr}
    \toprule
    Program & Basis     & $N_{\mathrm{P}}\,(N_{\mathrm{C}})$  & E [a.u.] &Error &$\Delta$E [a.u.]& Error & Time [s]\\
    \midrule
\multirow{4}{*}{LSDalton} & pc-1                           &2502 (1542)&-3770.20554 &2.6e-00 &-19.12076 &1.1e-01   &101     \\
                          & pc-2                           &5040 (3600)&-3772.56656 &3.0e-01 &-19.27598 &-4.0e-02  &1688    \\
                          & pc-3                           &9972 (8052)&-3772.83312 &3.3e-02 &-19.24894 &-1.4e-02  &26319   \\
                          & pc-3\textsuperscript{\emph{a}} &9972 (8052)&-3772.82592 &4.0e-02 &-19.26353 &-2.9e-02  &1924    \\
    \midrule                        
\multirow{3}{*}{ORCA} & pc-1&2502 (1542)&-3770.19956 &2.6e-00 &-19.11479 &1.2e-01   &25      \\
                      & pc-2&5040 (3600)&-3772.56922 &3.0e-01 &-19.27865 &-4.4e-02  &265     \\
                      & pc-3&9972 (8052)&-3772.83357 &3.2e-02 &-19.24940 &-1.4e-02  &3933    \\
    \midrule                    
\multirow{4}{*}{MRChem} & MW4 &&-3772.85028 &1.6e-02 &-19.21937 &1.5e-02   &117     \\
                        & MW5 &&-3772.86560 &2.5e-04 &-19.23469 &2.5e-04   &390$^*$  \\
                        & MW6 &&-3772.86584 &1.0e-05 &-19.23493 &1.0e-05   &1005$^*$ \\
                        & MW7 &&-3772.86585 &-       &-19.23494 &-         &2577$^*$ \\
    \bottomrule
    \end{tabular}    
    

\textsuperscript{\emph{a}} Performed using density fitting (df-def2 basis: $N_{\mathrm{P}}=9366$, $N_{\mathrm{C}}=7518$) and ADMM (admm-3 basis: $N_{\mathrm{P}}=4560$, $N_{\mathrm{C}}=3768$) to accelerate the construction of $J$ and $K$, respectively.\cite{Kumar2018-yc}
\end{table}

We notice that at the lowest precision (MW4), the results have
the same quality as the pc-3 basis, but at a fraction of the computational
cost. \ac{MW} calculations show also a rapid convergence pattern with
roughly a 2.5 times increase in computational time for each
additional digit in the precision. \ac{GTO} bases show instead a much
slower convergence trend: a three-fold error reduction has a cost of
more than an order of magnitude. Run times (and possibly the pc-4 convergence) for ORCA\cite{Neese2022-ub}
and LSDalton\cite{Aidas2014-rp} could likely be improved by tuning
different input parameters. We did not attempt to further optimize all
these settings: we are confident that the main picture would not
change.

We stress also that it would not be possible to assess the error of
\ac{GTO} calculations, were the precise \ac{MW} results not available.
Reducing the error by increasing the basis set size is computationally
demanding, if at all possible for large systems.
The ability to better control precision is a definitive advantage of the \ac{MRA} method:
the precision is determined by a single parameter that is independent on the property of interest and that can be chosen on a continuous scale.

\section{Discussion}

We have presented a fully functional implementation of the
multiwavelet method and demonstrated that it is capable of handling
systems with thousands of electrons at the \ac{HF}
level. The methodology is -- almost -- linear scaling with
system size. The precision can be adjusted as a single, user-defined input variable,
which is more intuitive compared to the choice of \ac{GTO} basis set
in traditional methods. For high precision or large systems, the
method is competitive with \ac{LCAO}
methods. It is only recently that computers large enough to treat such
systems at an affordable computational cost have been available; this
may explain why \ac{MRA} methods have received relatively less attention so
far.

There are certainly many alternative ways to approach the problem and
we still expect that significant improvements in the efficiency of the
algorithm will be implemented in the coming years: especially the
large memory footprint is a serious bottleneck that should be further
addressed.

The actual basis used in MRChem can be
several order of magnitudes larger than what is used in large \ac{GTO}
basis. Nevertheless, the computation times are still affordable and even
competitive. We believe this is due to modern computers being
extremely efficient at performing a large number of
mathematical operations simultaneously. Accessing and transferring data is by comparison
significantly more expensive. Moving data from main memory into
cache, and data transfer between compute nodes can affect
performance. The ability to
partition the problem into independent parts is becoming more
important than minimizing the overall number of mathematical
operations \cite{Gavini}. In this respect, the \ac{MRA} framework has an advantage
because of the orthogonality, by construction, of the basis. Similar points have been recently raised by other authors.
\cite{Penchoff2021-cl, Ratcliff2017-nr}
Algorithms that are designed by ``thinking parallel" are well-suited to effectively exploit the computational power of 
modern, massively parallel computing clusters.\cite{Mattson2004-oc,McCool2012-tx}

For low or medium precision, traditional basis set
approaches are still faster.  It is however fair to say that finite
basis set methods benefit from decades of development by thousands of
scientists, while only a few groups are currently developing
Multiwavelet methods for quantum chemistry calculations. We can
therefore expect that the potential for further optimizations of the
method and extension of the range of application will increase widely
in the future.

The current work has focused on the \ac{HF} method. For \ac{DFT}
methods, it has already been shown that
\ac{MRA} methods can be competitive with respect to computation
time.\cite{Bischoff2019-mr} The most demanding contribution, namely the exact exchange is
also present for hybrid \ac{DFT} functionals and the present
discussion therefore applies for \ac{DFT} without
substantial modifications. The remaining issues concern
correlated methods, which are still an active field of development for
\ac{MRA} methods.\cite{Bischoff2013, Kottmann2017, Kottmann2020}

\section{Code and data availability}
The MRChem code is available at \url{https://github.com/MRChemSoft/mrchem} \cite{mrchem}. The molecule geometries, main input parameters and model outputs can be found at \url{https://doi.org/10.18710/96NRIX} \cite{96NRIX_2022}.

\section{Author contributions}
\label{author_contrib}

We use the CRediT taxonomy of contributor roles.\cite{Allen:2014kj,Brand:2015jr}
The ``Investigation'' role also includes the ``Methodology'', ``Software'', and ``Validation'' roles.
The ``Analysis'' role also includes the ``Formal analysis'' and ``Visualization''
roles. The ``Funding acquisition'' role also includes the ``Resources'' role.
We visualize contributor roles in the following authorship attribution matrix, as suggested in Ref.~\citenum{authorship-attribution}.

\begin{table}
  \caption{Levels of contribution: \textcolor{blue!100}{major}, \textcolor{blue!25}{support}.}
  \label{tbl:example}
  \begin{tabular}{lccccccc}
    \toprule
                              & PW     & MB      & AB     & GAGS    & SRJ    & RDRE    & LF     \\
    \midrule
    Conceptualization         & \major &         &        &         & \major &         & \minor \\ 
    Investigation             & \major & \minor  & \minor & \minor  & \major & \minor  & \minor \\ 
    Data curation             & \major &         & \minor &         &        &         &        \\ 
    Analysis                  & \major &         &        &         & \minor & \minor  & \minor \\ 
    Supervision               & \minor &         &        &         & \minor &         & \minor \\ 
    Writing -- original draft & \major &         &        &         & \minor &         & \minor \\
    Writing -- revisions      & \major & \minor  & \minor & \minor  & \minor & \minor  & \minor \\
    Funding acquisition       &        &         &        &         &        &         & \major \\ 
    Project administration    & \major &         &        &         & \major &         & \major \\
    \bottomrule
  \end{tabular}
\end{table}

\begin{acknowledgement}
We acknowledge support from the Research Council of Norway through its Centres of Excellence scheme, project number 262695. 
The computations were performed on resources provided by Sigma2 - the National Infrastructure for High Performance Computing and 
Data Storage in Norway, through grant of computer time, no.
nn4654k and nn9330k.
R.D.R.E. acknowledges support from the European High-Performance Computing  Joint Undertaking under Grant Agreement No. 951732.
\end{acknowledgement}

\begin{suppinfo}
The supporting information file contains: a short introduction to the \ac{MRA} representation of functions using \ac{MW} bases, a detailed discussion of the mathematical operations needed to perform a \ac{SCF} cycle, and the available implementation strategies for each of them.
\end{suppinfo}

\bibliography{references}

\begin{figure}
\centering
\includegraphics[width=\textwidth]{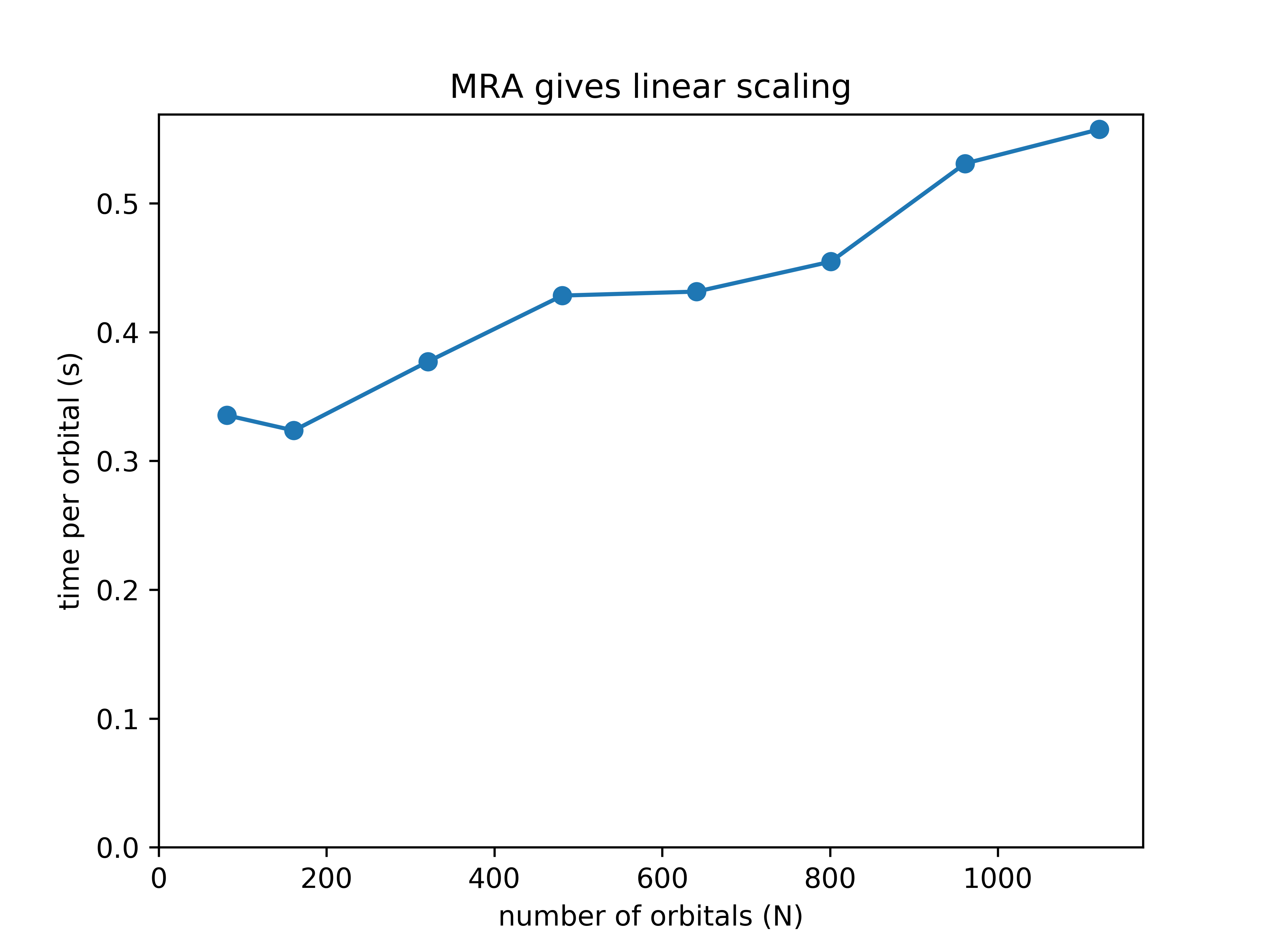}
\label{For Table of Contents Only}
\end{figure}

\end{document}


\maketitle

A note on this document: equations and sections referenced in the Supporting Information refer to the main manuscript, unless the reference number contains the SI prefix.

\section{Function representation with Multiwavelets}\label{sec:func_rep}

In order to make the implementation details clearer to the reader we will here introduce a few simple concepts of \ac{MRA}. We will by no means be exhaustive and refer the interested reader to the rich literature on the subject for further details.\cite{Alpert:2002cx,Frediani.10.1080/00268976.2013.810793,Beylkin.10.1002/cpa.3160440202,Bischoff2019-mr}. In particular, the 2002 article by \citeauthor{Alpert:2002cx} is a very pedagogical introduction to \acp{MW}.~\cite{Alpert:2002cx}

A one dimensional \ac{MRA} is obtained by defining a set of scaling functions $\scaling_p(x)$, which fulfill the so called self-similarity properties:
the $k+1$ scaling functions $(p=0,1 \ldots k)$ in interval $[l/2^s,(l+1)/2^s]$ can be obtained by translating and dilating the original functions defined in $[0,1]$:
\begin{equation}
    \scaling^s_{pl}(x) = 2^{s/2} \scaling_p(2^s x - l).
\end{equation}

The wavelet functions are  the orthogonal complement of the scaling functions between two successive scales $s$ and $s+1$ and they also fulfill a self-similarity relationship:
\begin{equation}
    \wavelet^s_{pl}(x) = 2^{s/2} \wavelet_p(2^s x - l).
\end{equation}

For details about the construction of the Multiwavelets used in this work we refer the reader to the original work of \citeauthor{Alpert.10.1137/0524016}. \cite{Alpert.10.1137/0524016}
For our purposes, it suffices to say that this construction provides a basis which is dense in $L^2$ and there are two equivalent ways to project a function, commonly called \emph{reconstructed} and \emph{compressed} representation. The reconstructed representation of a function $f$ makes use of the scaling functions at the finest scale $S$ (largest $s$ value), 
\begin{equation}\label{eq:reconstructed_func}
  f = \sum_{l=0}^{2^S-1}\sum_{p=0}^k c^S_{pl} \scaling^S_{pl}, \qquad c^s_{pl} = \braket{f | \scaling^s_{pl}}. 
\end{equation}
The compressed representation makes use of the scaling functions at $s=0$ and all successive wavelet functions from scale $s=0$ to scale $s=S-1$.
\begin{equation}\label{eq:compressed_func}
  f = \sum_{p=0}^k c_{p} \scaling_p + \sum_{s=0}^{S-1} \sum_{l=0}^{2^s-1} \sum_{p=0}^k d^s_{pl} \wavelet^s_{pl}, \qquad d^s_{pl} = \braket{f | \wavelet^s_{pl}}
\end{equation}
In practice, the \emph{reconstructed} representation describes the function as a piece-wise polynomial on a grid. Such a grid is generally \emph{adaptive} and the finest scale $S$ is therefore not a uniform value. The \emph{compressed} representation is instead used to determine the level of refinement which is necessary to keep the error under control.  Both function representations can be expressed as a tree structure where each pair of indices $sl$ designates a \ac{MW} node in the tree. The range of $l$ is both scale- and function-dependent: it runs over all nodes included at the given scale for the selected function/orbital. For the reconstructed representation the nodes in use are the \emph{leaf} nodes in the tree (nodes with no further ramifications), whereas for the compressed representation, the nodes in use are all the \emph{branch} nodes, starting from the single root node ($s=0$).
Switching from one representation to the other is performed through the fast \ac{MW} transform.\cite{Beylkin.10.1002/cpa.3160440202,Frediani.10.1080/00268976.2013.810793}. For the purpose of the present discussion, suffices to say that the \ac{MW} transform is local and fast: the computational cost of the \ac{MW} transform scales as the number of nodes in the function tree and requires only a few small matrix multiplications for each node.

For operator applications in the non-standard form\cite{Beylkin.10.1002/cpa.3160440202,Frediani.10.1080/00268976.2013.810793}, both scaling and wavelet coefficients are needed and they are therefore stored for each \ac{MW} node. As a result, each three dimensional node will contain some metadata (scale $s$, translation indices $l_x, l_y, l_z$, connectivity information about neighbors nodes, parent and children, etc...) and an array of $2^3 (k+1)^3$ coefficients: $(k+1)^3$ scaling coefficients and $(2^3-1)(k+1)^3$ wavelet coefficients.

\section{The basic operations in the SCF iterations}

We will here list the main operations involved in the \ac{SCF} procedure, with emphasis on those aspects which are relevant for an efficient parallelization. For a detailed description of the \ac{SCF} optimization using \acp{MW}, the reader is referred to a recently published article which describes the algorithm in detail.\cite{Jensen:2022gg}

The starting point is an initial set of orthonormal orbitals which define a Slater determinant for the given problem. A density is computed and the corresponding operators ($J$ and $K$) are built. The Fock matrix is computed as in Eq.~\eqref{eq:fock-matrix-definition}, which requires the application of the kinetic operator ($T$), nuclear potential ($V_{\mathrm{nuc}}$), Coulomb ($J$) and exchange ($K$) operators on each occupied orbital. The argument for the application of the integral operator is assembled (the square bracket in Eq.~\eqref{eq:integral-SCF}). Then the integral operator is applied separately for each orbital and the cycle is iterated after an orthonormalization step. Algorithm \ref{alg:SCF-iter} depicts these main steps.

\begin{algorithm*}
  \caption{Pseudocode for a single self-consistent field iteration. Each step takes advantage of adaptivity and screening, as described in Sections~\ref{sec:localvsorb} and~\ref{sec:adaptivity-screening}.} 
  \label{alg:SCF-iter}
  \begin{algorithmic}[1]
    \Procedure{SCF iteration}{$\lbrace \orbital_{i}^{[n]}\rbrace_{i=1}^{N_{\mathrm{occ}}}$, $\epsilon$}
    
    \State Build Coulomb operator: $$J^{[n]} =
    \sum_i \left[\frac{1}{|\rvec - \rvec^\prime|} \star\left(\orbital_{i}^{*,[n]}\orbital_{i}^{[n]}\right)\right]$$
    \Comment{See Section~\ref{sec:compute-J}}
    \State Build and apply Exchange operator: $$K^{[n]}\orbital_{j}^{[n]} = \sum_i \orbital_{i}^{[n]} \left[\frac{1}{|\rvec - \rvec^\prime|} \star\left(\orbital_{i}^{*,[n]}\orbital_{j}^{[n]}\right)\right]$$
    \Comment{See Section~\ref{sec:compute-K}}
    \State Assemble Fock matrix elements: $$F_{ij}^{[n]} = \Braket{\orbital_{i}^{[n]} | T + V_{\mathrm{nuc}} + J^{[n]} - K^{[n]}| \orbital_{j}^{[n]}}$$
    \Comment{See Sections~\ref{sec:compute-T} and~\ref{sec:compute-V}}
    \State Assemble right-hand side: $$V^{[n]}\orbital_{i}^{[n]} - \sum_{j\neq i}^{N_{\mathrm{occ}}}F_{ij}^{[n]}\orbital_{j}^{[n]}$$
    \Comment{See Section~\ref{sec:rhs}}
    \State Update orbitals: $$\orbital_{i}^{[n+1]} = -2H^{\mu_{i}}\star\left[ V^{[n]}\orbital_{i}^{[n]} - \sum_{j\neq i}^{N_{\mathrm{occ}}}F_{ij}^{[n]}\orbital_{j}^{[n]}\right]$$
    \Comment{See Section~\ref{sec:rhs}}
    \State Accelerator and localization: A new set of orbitals is created using results from previous iterations and localization. This step is optional and does not need to be performed at each iteration.
    \EndProcedure
  \end{algorithmic}
\end{algorithm*}

The elementary mathematical operations required in each SCF iteration are:
\begin{itemize}
\item Projection of functions in a \ac{MW} representation, given their analytic form or a representation on an arbitrary grid.\footnote{As long as a function is \emph{algorithmically} known, that is, there exists an implementation that given a point in the domain, for example $\mathbb{R}^{3}$, returns the function value, then it is possible to obtain its \ac{MW} representation.} 
\item Evaluation of the \ac{MW} representation of functions at a given point in space.
\item Multiplication of a function by a number.
\item Sum of two \ac{MW} functions.
\item Product of two \ac{MW} functions.
\item Computation of the overlap integral between two \ac{MW} functions.
\item Application of a multiplicative operator, \emph{e.g.} a potential, on a \ac{MW} function.
\item Evaluation of the first order derivative of a \ac{MW} function.
\item Construction of the separated representation of the Poisson and Helmholtz kernels as \ac{MW} functions.
\item Application of the Poisson or Helmholtz operators on a \ac{MW} function.
\end{itemize}

All these operations are provided by the MRCPP library\cite{mrcpp}. They must be composed to achieve a \ac{SCF} implementation, and they need to be integrated in a parallel framework to achieve good scaling and performance. In addition, the \ac{SCF} cycle requires schemes for orbital orthonormalization, rotation, and localization, Fock matrix diagonalization, history storage and utilization to enable convergence acceleration, using \emph{e.g.}~the \ac{KAIN} method.~\cite{Harrison2004-fs}

\subsection{The kinetic operator $T$}
\label{sec:compute-T}
The application of the kinetic energy operator on a \ac{MW} function is formally not well defined due to the discontinuities in the basis.\cite{Anderson2019-bx} However, when adopting an integral reformulation of the \ac{SCF} equations, the kinetic energy operator is required only in the  expectation value expressions as in Eq.~\eqref{eq:fock-matrix-definition}, and as shown by \citeauthor{Anderson2019-bx}~\cite{Anderson2019-bx} the original definition of the first-order derivative by \citeauthor{Alpert:2002cx}~\cite{Alpert:2002cx} is able to deliver sufficient precision in this specific case. On the basis of this consideration, and for computational efficiency, the matrix elements of the Laplacian operator are computed as inner products of orbital gradients:
\begin{equation}
  \label{eq:kinetic-matrix}
  T_{ij} = 
  -\frac{1}{2} 
  \Braket{\orbital_i | \nabla^2 \orbital_j} 
  = 
  \frac{1}{2} 
  \Braket{\nabla \orbital_i | \nabla \orbital_j}.
\end{equation}

It is possible to avoid the kinetic energy operator altogether \cite{Jensen:2022gg}, but in practice this complicates the algorithm to the point where the direct computation as in Eq.~\eqref{eq:kinetic-matrix} is to be preferred for large systems.

\subsection{The nuclear potential $V_{\mathrm{nuc}}$}
\label{sec:compute-V}
The nuclear potential consists of a sum of contributions from the different nuclei. As such, applying it to a given orbital can be performed in two ways.

\begin{description}
\item[apply-then-sum] The potential from each nucleus $\alpha$ is applied to each orbital and the results summed up over all nuclei:
\begin{equation}
  V_{\mathrm{nuc}} \orbital_j = \sum_{\alpha}^{N_{\mathrm{atoms}}} \left(V_{\alpha} \orbital_j\right),
\end{equation} 
\item[sum-then-apply] The sum over the nuclei is performed first, followed by the multiplication with the given orbital:
\begin{equation}
V_{\mathrm{nuc}} \orbital_j = \left(\sum_{\alpha}^{N_{\mathrm{atoms}}} V_{\alpha}\right) \orbital_j.
\end{equation}
\end{description}

The first approach requires $N_{\mathrm{atoms}}\times N_{\mathrm{occ}}$ function multiplications, followed by a summation of $N_{\mathrm{atoms}}$ terms. It also needs memory for $N_{\mathrm{atoms}}\times N_{\mathrm{occ}}$ \ac{MW} functions. However, since these objects are relatively localized, their memory footprint is modest and can, in addition, be amortized by distributing objects across parallel processes. Furthermore, the function-function products can also be distributed across nodes and screened aggressively.

The second approach requires a summation of $N_{\mathrm{atoms}}$ terms, followed by $N_{\mathrm{occ}}$ function multiplications. However, the total potential is no longer well-localized, as it describes the potential from all clamped nuclei, and it requires a large amount of memory storage. Storing the nuclear potential is wasteful: only its \ac{MW} nodes with non-zero overlap with the occupied orbitals will contribute to the final result.

Despite these considerations, our present implementation uses the ``sum-then-apply" form. In the future we plan to try the first approach and also adapt the accuracy according to the electronic density and compare the performance between the two approaches.

\subsection{The Coulomb potential $J$}\label{sec:compute-J}
The Coulomb interaction can be expressed as:
\begin{equation}
    \begin{aligned}
  J(\rvec) &= \sum_i \int \frac{\orbital^*_i(\rvec')\orbital_i(\rvec')}{|\rvec-\rvec'|} \, \mathrm{d}\rvec' =
  \sum_i  \left[
  \frac{1}{|\rvec-\rvec'|} \star 
  \orbital^*_i\orbital_i
  \right] =
  \frac{1}{|\rvec-\rvec'|} \star 
  \left[\sum_i 
  \orbital^*_i\orbital_i
  \right]
    \end{aligned}
\end{equation}
where, as indicated by the last step, we can choose the order of the operations when constructing the Coulomb potential: either summing the orbital densities first and compute the potential of the total density (\textbf{sum-then-apply}), or, first computing the Coulomb potential of each individual orbital density and then sum the potentials (\textbf{apply-then-sum}). The electrostatic potential is in general a much smoother function than its corresponding density, which contains cusps at every nuclear site. It is thus desirable to avoid the explicit construction of the global density and instead gradually build a global potential by partial applications of the Poisson operator to subsets of orbital densities. This operation can also easily be distributed in parallel by assigning different subsets to different processes.

The Poisson kernel is expressed in a separable form, within the requested precision, as a sum of Gaussian contributions.\cite{Beylkin.10.1016/j.acha.2005.01.003} Each Gaussian kernel is applied separately and the contributions are then added up. The separated representation is essential to achieve fast linear-scaling algorithms: on the one hand it brings down the formal scaling of the operator application for each node from $2^6 k^6$ to $M 2^4 k^4$, where $k$ is the polynomial order and $M$ is the separation rank; on the other hand the Gaussian contributions will range from wide ones (shallow \ac{MW} representation which couple long-range but at coarse scales) to narrow ones (deep \ac{MW} representation which couple short-range at all scales). In practice, most of the $M$ terms will be applied very efficiently and the separation between long-range and short-range interactions -- which is otherwise achieved through the \ac{FMM} -- is naturally included without extra effort.\cite{Frediani.10.1080/00268976.2013.810793, Beylkin.10.1016/j.acha.2007.01.001}

Normally the potential is computed with the required precision in the entire space, and since it is long range, it will be relatively memory consuming.

\subsection{The exchange potential $K$}\label{sec:compute-K}
The construction and application of the exchange potential is the most computationally demanding part, because the exchange operator cannot be written as a simple multiplicative potential. A Coulomb potential must be constructed and applied for each orbital pair:

\begin{equation}
  K[\orbital_j] = \sum_i K_i[\orbital_j] = 
  \sum_i \orbital_i(\rvec) \int \frac{\orbital^*_i(\rvec')\orbital_j(\rvec')}{|\rvec-\rvec'|} \, \mathrm{d}\rvec'
  =
  \sum_i \orbital_i(\rvec) \left[
  \frac{1}{|\rvec-\rvec'|} \star 
  \orbital^*_i\orbital_j
  \right]
\end{equation}
Section \ref{Xscreen} provides a more detailed description of how these terms can be evaluated efficiently in a \ac{MRA} framework. 

\subsection{Construction of the right-hand side and Helmholtz operator application}\label{sec:rhs}
The central step in the SCF iteration is the application of the Helmholtz operator $H^{\mu}$ defined in Eq.~\eqref{eq:helmholtz-green} to the square bracket on the right-hand side in Eq.~\eqref{eq:integral-SCF}. The Helmholtz kernel depends on the orbital energy ($\mu_i = \sqrt{-2F_{ii}}$). As such, one needs to construct as many operators as there are occupied orbitals. Given the Fock matrix, each term within the square bracket in Eq.~\eqref{eq:integral-SCF} can be assembled with $O(N)$ formal operation count per orbital $\orbital_i$. Thus, the formation of the whole right-hand side for the entire set of orbitals formally scales as $O(N^{2})$. However, $\sum_{j \ne i}^{N_{\mathrm{occ}}} F_{ij} \orbital_j $ is a linear combination of orbitals and, as shown in Section~\ref{sec:localvsorb}, it can be performed efficiently with $O(N)$ operation count. Also the application of the exchange operator can be done essentially in $O(N)$ operations as shown in Sections \ref{Xscreen} and \ref{sec:screen}.

Once the right-hand side is assembled, the application of the Helmholtz operator is practically identical to the Coulomb operator. The only difference is the actual expansion in Gaussians, which, thanks to the long-range exponential decay of the kernel, has more favorable scaling properties.\cite{Frediani.10.1080/00268976.2013.810793} 

\bibliography{references}